\documentclass[conference]{IEEEtran}
\IEEEoverridecommandlockouts
\usepackage{cite}
\usepackage{amsmath,amssymb,amsfonts}
\usepackage{algorithmic}
\usepackage{graphicx}
\usepackage{textcomp}
\usepackage{pgfplots}
\usepackage{xcolor}
\usepackage{booktabs}
\usepackage{balance}
\def\BibTeX{{\rm B\kern-.05em{\sc i\kern-.025em b}\kern-.08em
    T\kern-.1667em\lower.7ex\hbox{E}\kern-.125emX}}

\DeclareRobustCommand*{\IEEEauthorrefmark}[1]{%
  \raisebox{0pt}[0pt][0pt]{\textsuperscript{\footnotesize #1}}%
}

\newcommand{\solution}{\textit{MTFS}}

\newcommand{\ie}{\textit{i.e., }}  % from the Latin id est = that is
\newcommand{\eg}{\textit{e.g., }}  % from the Latin exempli gratia = for example
\def \1{\textit{(i)}}
\def \2{\textit{(ii)}}
\def \3{\textit{(iii)}}
\def \4{\textit{(iv)}}
\def \5{\textit{(v)}}

\begin{document}
\author{
    \IEEEauthorblockN{Jan von der Assen\IEEEauthorrefmark{1}, Alberto Huertas Celdrán\IEEEauthorrefmark{1}, Rinor Sefa\IEEEauthorrefmark{1}, G\'er\^ome Bovet\IEEEauthorrefmark{2}, Burkhard Stiller\IEEEauthorrefmark{1}}
    
    \IEEEauthorblockA{\IEEEauthorrefmark{1}Communication Systems Group CSG, Department of Informatics, University of Zurich UZH, CH--8050 Zürich, Switzerland \\{[vonderassen, huertas, stiller]}@ifi.uzh.ch, rinor.sefa@uzh.ch}
    
    \IEEEauthorblockA{\IEEEauthorrefmark{2}Cyber-Defence Campus, armasuisse Science \& Technology, CH--3602 Thun, Switzerland gerome.bovet@armasuisse.ch}
    
%    \IEEEauthorblockA{\IEEEauthorrefmark{3}Department of Information and Communications Engineering, University of Murcia, 30100--Murcia, Spain {gregorio}@um.es}
    
}

\title{\solution{}: a Moving Target Defense-Enabled\\  File System for Malware Mitigation}

\maketitle

\begin{abstract}
Ransomware has remained one of the most notorious threats in the cybersecurity field, for which Moving Target Defense (MTD) has been proposed as a novel defense paradigm. Although various approaches leverage MTD, few of them rely on the operating system and, specifically, the file system, thereby making them dependent on other computing devices, rendering defense against certain threats unrealistic. File-based approaches are less studied here, while showing limitations in resource usage and defense effectiveness. Furthermore, existing ransomware defenses merely restore data or detect attacks, without preventing them. %The \solution{} approach presented in this paper enables MTD from the file system perspective. 
Thus, this paper introduces the \solution{} file system and the design and implementation of three novel MTD techniques -- one delaying attackers, one trapping recursive directory traversal, and another one hiding file types. The effectiveness of the techniques are shown in three experiments. First, it is demonstrated that the techniques can delay and mitigate ransomware on real IoT devices. Secondly, in a broader scope, the solution was confronted with 13 ransomware samples, highlighting that it can save 97\% of the files. In terms of overhead, the defense system consumes only a small amount of resources, highlighting the feasibility of proactive defense.

\end{abstract}


\section{Introduction}
\label{intro}

Malware is one of the most prominent attack vectors in the still highly active threat landscape that today's enterprises find themselves in. The frequency at which malware-based attacks are launched has increased by 358\% in 2020~\cite{malware-358}. Among malware-based attacks, ransomware is a frequent threat, leading to devastating impacts. In 2022, 20\% of all cybercrime were attributed to ransomware attacks, where the average cost ranges between 1 and 8 Million USD~\cite{malware-358}. %This increment is partially motivated by Internet-of-Things (IoT) devices, which are resource-constrained computing devices that are especially vulnerable to attacks due to missing or improper security controls.

Fixing the vulnerabilities that are exploited in security breaches is touted as the most important measure for prevention. However, the reality with respect to how devices are administrated draws a dramatically different picture, with security patches either not being targeted, developed, or deployed. With respect to ransomware, it was observed that many attacks exploit age-old system flaws~\cite{vulnerabilities}. All of this magnifies the importance of a defense in depth approach, which includes multiple security controls. For example, an industrial IoT device may not allow the modification of the firmware by the operator without voiding the warranty. In this case, there is no option to fix the vulnerability, and the only way to improve security is to add additional defense layers.

One promising paradigm to develop additional security controls is Moving Target Defense (MTD)~\cite{mtd-metrics}. By creating dynamicity (\ie moving) in the elements comprising the attack surface (\ie the target), the complexity for the attacker is increased, thereby decreasing the likelihood of a successful attack (\ie the defense). As identified by a recent survey~\cite{mtd-where-art-thou}, MTD for IoT is a promising yet immature defense paradigm. The key limitations identified include the lack of solutions that employ MTD at the operating system (OS) level, since most solutions rely on network-based approaches, which may not be applicable for every threat. %, a dependency that may not be applicable for every defense and threat model. 
%For example, a ransomware attack may not be mitigated from the network point of view, once encryption keys are delivered and the encryption is ongoing. 
As identified through a literature review, MTD approaches in this field did not explore the applicability of a purely virtualized file-based approach for MTD to increase the  performance efficiency and defense effectiveness.  %delivered with a thumb drive can not be detected before the damage is already done.
Furthermore, existing MTD approaches are considered to be of limited maturity, since only few approaches have actually been implemented and tested in real scenarios.

To cover the previous challenges, this paper introduces \textit{MTFS}, a file system specifically created as a platform to home MTD techniques on the OS-level. Secondly, to present the effectiveness of \solution{}, three deceptive MTD techniques (\ie combating directory traversal, file access, and file type identification) are implemented into \solution{} to mitigate ransomware breaches in Linux devices. %To show that the file system and the mitigation techniques are able to run on an IoT device, this paper evaluates the defense layer by deploying all components in a production ElectroSense sensor and then infecting the device with a set of ransomware samples.  
The defense layer provided by \solution{} and the three proposed MTD techniques have been evaluated in a real resource-constrained device affected by a ransomware. The selected device is a Raspberry Pi 3, acting as a radio frequency spectrum sensor. To highlight the portability of the file system to other Linux devices, while remaining effective, a container-based testbed is created to evaluate the system. With this testbed, \solution{} is evaluated against several ransomware samples. The experiments conducted in the IoT device and the testbed showed that the \solution{}-based MTD approach successfully operates on a limited number of resources, that the defense model holds up in a realistic deployment scenario, and that it mitigates numerous ransomware samples.

The remainder of the paper is structured as follows. Section~\ref{rw} introduces the MTD field and highlights the lack of works in the described direction. Section~\ref{sol} introduces the \solution{} file system and the prototypical mitigation techniques, which are evaluated in three experiments presented in Section~\ref{eval}. Finally, Section~\ref{conc} summarizes the findings. % and draws conclusions from the results.

\section{Background and Related Work}
\label{rw}
\setlength{\tabcolsep}{5pt}
\begin{table}[b]
\centering
\caption{Related Moving Target Defense Approaches}
\begin{tabular}{@{}llllll@{}}
\toprule
            \textit{Sol.}         & \textit{Attack}  & \textit{What?} &  \textit{How?}&   \textit{When?} & \textit{Impl.} \\ \midrule
           \cite{jajodia_introducing_2011} 2011 & Unknown  & Software  & Divers. & Random   & Idea \\ 
           \cite{butts_creating_2011} 2011 & Unknown   &  HW, OS                      &   Divers.                    &     Events                     &     Live                           \\ 
           \cite{jajodia_toward_2011}  2011 & DoS  & Network                      &  Shuff.                     &     Interval                     &   Idea                             \\ 
           \cite{vikram_nomad_2013} 2013  & Botnets      & HTML          & Shuff.             & Events    & Live                           \\ 
           \cite{gomez-boix_collaborative_2019} 2019 & Fingerprinting   & Browser                 & Shuff.                      &  Events                         & Idea       \\
           \cite{roy_moving_2019}  2019 &  Input &   Algorithm & Divers. & Events & Live \\
           \cite{abdelnabi_whats_2021} 2021 &  Input  & Model & Divers. & Events & Live             \\
           \cite{voulimeneas_distributed_2020} 2020 & Memory  & Deployment & Redund.  & Events   & Live   \\
           \cite{thompson_multiple_2014} 2014 & Reconn.   & OS               &  Divers.                      &  Interval                         & Live                               \\
           \cite{brown_dynamic_2020} 2020 & Reconn.            &  CAN bus               &  Shuff.            &  Proact.     & Virt.            \\ 
           \cite{duan_range_2020} 2020 & Reconn. &  Access Point & Redund. & Events & Idea \\
           \cite{icc} 2023 & Ransomware & Files & Shuff. & Events & Live  \\
           \textit{This} & Ransomware & Virtual Files & Shuff. & Events & Live  \\
           \midrule
\end{tabular}%
\label{table:MovingTargetDefenseRelatedWorkTable}
\end{table}

Moving Target Defense (MTD) aims to decrease the attack success probability  by making the attack surface more dynamic. The underlying principle tackles the attack on static targets, which exhibits an asymmetry where attackers have all the time they need to attack a target. %In addition, attackers only need to find one exploitable vulnerability whereas defenders need to ensure that every system flaw is met with a security control, which in turn should not exhibit security flaws.
A recent review of the MTD field for IoT devices revealed a lack of approaches that do not rely on third-party actors while presenting a mature implementation and evaluation~\cite{mtd-where-art-thou}. Aside from concluding on the immaturity of existing approaches, this survey introduced the key design decisions of \textit{What}, \textit{How}, and \textit{When} to apply a transformation to the attack surface. \textit{What} refers to the element of the attack surface that will be changed dynamically. \textit{How} defines a transition strategy (\ie making it more diverse, shuffling between sets of parameters, or increasing redundancy) which can be invoked based on a specific interval  or event, which constitutes the \textit{When}. As highlighted in \tablename~\ref{table:MovingTargetDefenseRelatedWorkTable}, numerous approaches have been presented that all focus on specific attack types, while defining \textit{What}, \textit{When}, and \textit{How} to mitigate the attack. For all of these approaches, the  implementation is crucial to understand their applicability in a real-world setting.

From the attack vector perspective, \cite{jajodia_introducing_2011, butts_creating_2011} introduce diversification-based approaches that aim to increase the defense capabilities without specifying a specific attack vector. Interestingly, they are implemented in different layers of the defense host. While \cite{jajodia_introducing_2011} focuses on the software stack in the browser, \cite{butts_creating_2011} aims to implement execution platform diversity on the hardware and OS-level.
\cite{jajodia_toward_2011, vikram_nomad_2013, gomez-boix_collaborative_2019} all rely on a shuffling strategy. They tackle different attack types such as Denial-of-Service, Botnet, and browser-based ones, but without implementing an approach in the operating system. 
Application-layer MTD approaches that rely on event-based diversification are \cite{roy_moving_2019, abdelnabi_whats_2021}. As such, they both address common threat vectors launched against machine learning based systems. Finally, \cite{thompson_multiple_2014, brown_dynamic_2020, duan_range_2020} present approaches to combat reconnaissance attacks, while leveraging specific network devices, such as Access Points and Controller Area Network Buses. Out of them, only \cite{thompson_multiple_2014} relies on the operating system. In \cite{thompson_multiple_2014}, the whole operating system (or distributions of the same operating system) is exchanged to avoid attacks. Most closely related to the work at hand, \cite{icc} presents an MTD framework that demonstrated how ransomware can be mitigated by shuffling between a set of (real) files, that are created on the application-level in user space.

\begin{table}[h]
\caption{File System Protection Solutions Against Ransomware}
\resizebox{\columnwidth}{!}{%
\begin{tabular}{@{}llllll@{}}
\toprule
            \textit{Solution}         &  \textit{Detection} &  \textit{Prevention} & \textit{Recovery}
            \\ \midrule
            \cite{continella_shieldfs_nodate}, 2016 & I/O classification  & None & Replication\\

            \cite{giuffrida_hidden_2018}, 2018 & Decoy objects & Security domains & None \\
            \cite{matos_rockfs_2018}, 2018 & None & None & Log System \\
            \cite{mcintosh_enforcing_2021}, 2021 & File-type validation & None & None \\
            \cite{lee_rcryptect_2022}, 2022 & Entropy analysis  & Authorization & None \\
           \midrule
\end{tabular}%
}
\label{table:FileSystemRelatedWorkTable}
\end{table}

Beyond the field of MTD, the advantages of a file system approach have been explored more clearly. As shown in \tablename~\ref{table:FileSystemRelatedWorkTable}, a number of papers explored the suitability of using a file system as a security measure. More specifically, related work in this area demonstrates that cyberattacks, such as ransomware, can be efficiently detected from the file system perspective. Furthermore, some approaches combine this detection with a file recovery component. This enforces the overall view of file system-based approaches, posing an opportunity to implement MTD. As such, related approaches did not explore proactive defense mechanisms against malware, which is the motivation of this work.  %In that sense, it can be demonstrated that it is possible to also actively and dynamically mitigate these attacks and not just detect and recover their effects.

In summary, related work in the MTD field, does not cover approaches leveraging the file system. In the broader sense, OS-level implementations like \cite{thompson_multiple_2014} and \cite{butts_creating_2011} have explored the same abstraction level and trust model as this work. However, there are no MTD approaches implemented in the file system-abstraction of an OS, although approaches such as~\cite{icc} show the applicability of using (real) files to combat malware. Since approaches outside the MTD paradigm have shown that the file system is a viable anchor to implement security mechanisms, there is an opportunity to explore this aspect of the operating system for dynamic, proactive defense. This is critical, since the operating system is, in many scenarios, the only dependable area of attack surface to implement MTD without relying on other actors, that may not be aligned with the threat and defense model.
\section{The MTFS Approach}\label{sol}
%This paper tries to solve the problem of detecting and mitigating ransomware attacks as close to the target asset (\ie the data held by the files) without losing the semantics (\eg calling process, operation type) of the attack behavior. For example, implementing such an approach on the block storage abstraction would ignore the semantics of the file abstraction.  %As such, this work considers the applicability of a file-system-based approach to implement dynamic and proactive defense following the MTD paradigm. In addition, a resource-constrained execution environment is considered to implement the approach.

\begin{figure}
    \centering
    \includegraphics[trim={.7cm 1.4cm .87cm .7cm},clip, width=\linewidth]{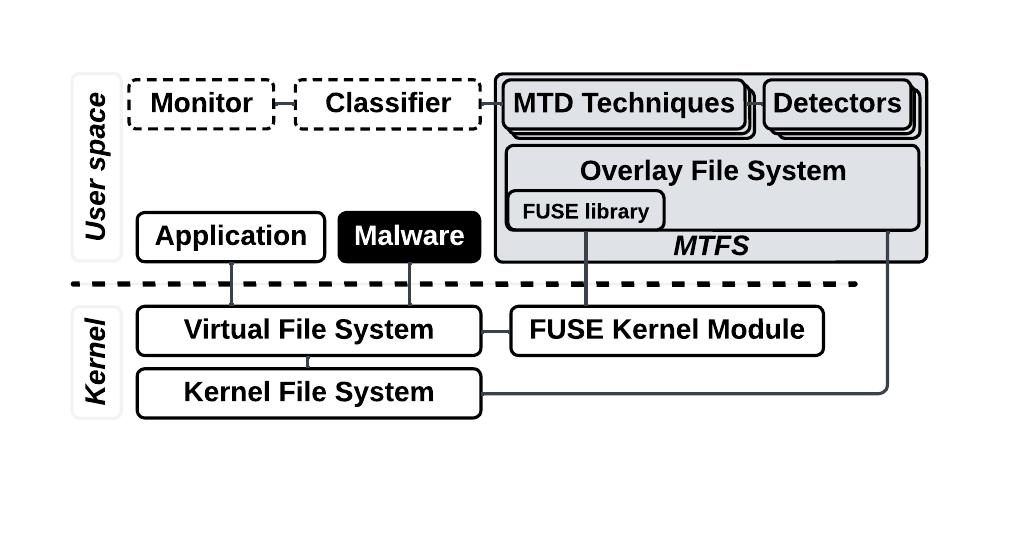}
    \caption{User Space Implementation of \solution{}}
    \label{fig:architecture}
\end{figure}

To achieve defense as close to the target asset (\ie file system data) as possible, \solution{} stands as a novel file system specifically created to detect, prevent or mitigate malware-based attacks. The file system abstraction is selected as it still preserves file operation semantics. \figurename~\ref{fig:architecture} presents the architectural elements that underly \solution{}, which follows the Linux file system architecture. The key elements added by \solution{} are a set of \textit{MTD techniques} which can be deployed based on \textit{Detectors}, which are all implement in an \textit{Overlay File System} that uses the \textit{FUSE} library.

To enable the interaction between these components, \solution{} makes use of the FUSE library to receive and respond to system calls, enabling full visibility into all operations carried out on the mount point of the file system. In addition, fine-grained control over how a process is being serviced can be implemented. This is advantageous for multiple reasons. First, this allows analysis of the call, before executing it. Secondly, it allows that a file operation is controlled to decide how to proceed with the call. In contrast to other file-based approaches, damage can theoretically be prevented before it is done. As such, \solution{} supports the integration of a set of \textit{Detectors} and \textit{MTD techniques}. The former are simply a set of strategies that receive for each process the system call specifications. These components can then evaluate in a bespoke manner how to analyze the call, for example, by evaluating them against a set of policies or by classifying them using a Machine-Learning model. \textit{MTD techniques}, on the other hand, are strategies that define specific actions to be invoked. With the architecture presented herein, they can also be invoked by outside components. This is especially useful when considering the plethora of existing classifiers and anomaly detectors that can provide insight from other sources (\eg performance metrics, other system calls). To achieve this integration and to provide users with a simple user space implementation, \solution{} uses the FUSE library, which provides client libraries in many high-level programming languages (\eg Python, Java, Go).

The remainder of the architecture consists of the Linux file system abstraction. Here, all user space applications, including malicious ones like ransomware, access data through the virtual file system, which is an abstraction present in the Linux kernel. Essentially, this layer abstracts the complexity of different file systems and storage media. Communication between user and kernel space is achieved with system calls, that allow operations (\eg \texttt{OPEN}, \texttt{CLOSE}, \texttt{STAT}, \texttt{READ}, \texttt{WRITE}) to be carried out. Based on the files referenced by the file descriptors, the operations can be routed to the underlying implementation of the file system (\eg ext4, FAT, FUSE). These implementations then provide access to the actual data. %To implement \solution{}, FUSE, was chosen as a reference architecture to implement the \solution{} subsystem in the file system. With FUSE, a kernel module is inserted into the kernel, which is connected to a file system implementation running in user space. 

%\subsection{Prototype Implementation}
To demonstrate the feasibility of the proposed file system-based approach and the effectiveness of building a security control within the file system, \solution{} was implemented as an overlay file system. Furthermore, three novel MTD techniques were designed and implemented on top of \solution{} to enable experiments with real ransomware samples. The overlay file system is a stacked file system that proxies requests to any underlying file system. %\textit{Ext4} is the default file system in the majority of Linux distributions, however, any file system works as an underlay file system. The overlay layer is a file system that forwards both the request and the response from the base file system to the  process. 
In addition to forwarding requests, the three MTD techniques represent the potential modifications that can be carried out in-place. The advantage of an overlay file system is that all operations can be intercepted and modified at will without inheriting the complexity of a native file system. To implement a prototype, the \texttt{go-fuse} library written in the \texttt{Go} programming language was used. 

The first of the integrated techniques \texttt{MTD\_DELAY} presents the simplest deception mechanism. Here, any requests to the overlay file system are serviced by the underlying file system. However, different delays can be specified to delay or hinder the attacker. Naturally, this does not completely prevent any attacks by itself. However, combined with another detection or prevention system, it can extend the time needed for defense. The mean time to attack (MTTA), is decreased, which is an important metric when discussing MTD techniques~\cite{mtd-metrics}.

The second MTD technique \texttt{MTD\_INF} presents malware samples with an \textit{Infinite Directory Graph}. This is based on the analysis of multiple ransomware samples, which all conduct a depth-first search to recursively find all files in a directory to be encrypted. Here, the technique shuffles between a set of real files and a self-linking directory. If a malware sample invokes the \texttt{opendir()} command, the technique responds with a subset of the underlay file system and a special directory names \texttt{!}, which is the first letter of the ASCII table. Invoking \texttt{opendir()} on this directory, in turn, yields the directory itself. Thus, conducting a depth-first search on this directory leads to a vicious circle, potentially disrupting or even preventing the encrypting process.

The third and final technique \texttt{MTD\_SUFFIX} is the most elaborated since it inspects responses from the underlay file system and modifies certain patterns. This technique is based on the observation that many ransomware samples adapt their behavior to the file type. \cite{icc} demonstrated that it is possible to prevent encryption by changing file suffixes. However, this approach has two limitations. First, changing entries in the real file system is a costly operation. Secondly, malicious software can still detect the type of file by looking at the first bytes (\ie magic numbers) of a file. This is where the third technique intercepts the \texttt{read()} system call to randomize this signature. Ransomware, relying on the system to work to achieve encryption, must therefore ignore unknown file types, focusing only on known file types (\eg plain text, PDF files).

\begin{figure*}
    \centering
    \includegraphics[trim={0.8cm 0.58cm 0.77cm 0.55cm},clip,width=0.9\linewidth]{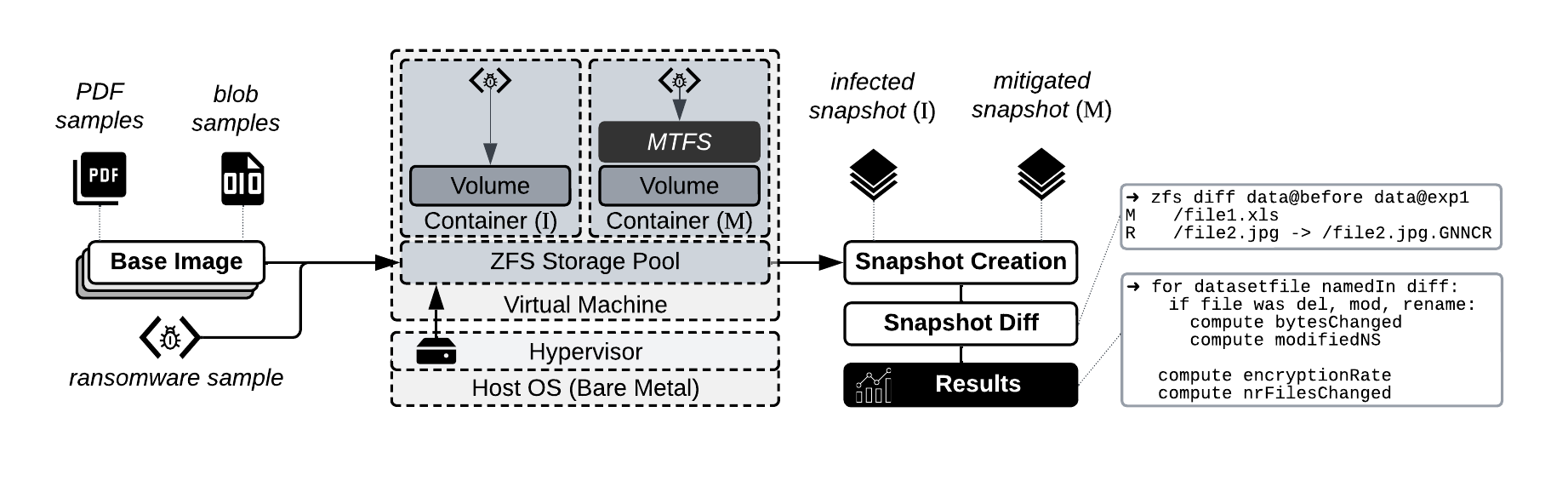}
    \caption{Architecture for the Ransomware Testbed}
    \label{fig:testbed}
\end{figure*}

The three MTD techniques described here do not implement an intelligence component by themselves, they expect that the malware sample can be detected on the device. Nevertheless, all techniques can be deployed proactively (\ie without an event). Still, an existing ML-based classification from~\cite{icc} was integrated to conduct experiments with the file system. This classification system monitors the host device's behavior with respect to resource consumption, expressed by roughly 80 parameters collected from different event families such as network, memory, file system, CPU, process scheduler, or device drivers~\cite{icc}.

\section{Evaluation}\label{eval}
\solution{} was assessed in three of experiments focusing on ransomware. First, the performance of the platform was evaluated by deploying it in a real sensor of the ElectroSense crowdsensing platform. Secondly, \solution{} was assessed by crafting a tailored testbed to evaluate several available samples and, to analyze the performance overhead.

\subsection{Real-World Scenario}
The real-world scenario targets a Raspberry Pi 4 Model B Rev 1.4 running the official ElectroSense image, which enables the data collection from the radio equipment~\cite{icc}. The Raspberry Pi board includes a 64 GB class 10 microSD card with a performance rating of 10. A subset of 1000 files from the govdocs~\cite{garfinkel2009bringing} dataset were deployed in the home directory of the device. With \textit{RansomwarePoC} and \textit{DarkRadiation}, two ransomware samples were used to test in the real environment. For each sample, the encryption performance in the real file system and in each of the three MTD techniques was assessed, as shown in \tablename~\ref{table:EvaluationResults}. Without any protection, all files are encrypted by the two samples, although \textit{DarkRadiation} is substantially faster. When mounting the file system, \textit{RansomwarePoC} was not able to encrypt any files in all the three cases. For the latter two strategies, the samples crashes, while for the first one it spends significant time trying to traverse the file system. \textit{DarkRadiation} was able to encrypt files in the delaying strategy, although the encryption duration was prolonged. For the other strategies, no files were lost.
\begin{table}[b]{%
\caption{File System Protection Effectiveness Against Ransomware}
\label{table:EvaluationResults}
\begin{tabular}{@{}lp{3.5cm}llll@{}}
\toprule    
            \textit{Ransomware}  
            & \textit{Implementation} &  \textit{Lost files} &  \textit{Time} 
            \\ \midrule
            RansomPoc & Default file system & 946/979 & 7m 51s \\
             & \texttt{MTD\_INF} & 0 & 10m 32s \\
            & \texttt{MTD\_SUFFIX} & 0 & 4s \\
            & \texttt{MTD\_DELAY} & 0 & 4s \\
           \midrule
           DarkRadiation & Default file system & 976/979 & 1m 14s \\
             & \texttt{MTD\_INF} & 0 & 31m \\
            & \texttt{MTD\_SUFFIX} & 0 & 1m 50s \\
            & \texttt{MTD\_DELAY} & 976/979 & 15m 7s \\
           \midrule
\end{tabular}%
}
\end{table}

\begin{figure*}
    \centering
        \begin{tikzpicture}
        \begin{axis}[
            width=\linewidth,
            height=4.5cm,
            %small,
            ybar,%=8pt, % configures ‘bar shift’
            ylabel={Changes [MB]},
            ymajorgrids=true,
            scaled ticks=false,
            symbolic x coords={
              Babuk, Blackbasta, Buhti, Cl0p, Conti,  Darkradiation,   GonnaCry, JavaRansomare, Lockbit, lollocker, monti, raasnet, RansomarePoC},
            xtick={
              Babuk, Blackbasta, Buhti, Cl0p, Conti,  Darkradiation,   GonnaCry, JavaRansomare, Lockbit, lollocker, monti, raasnet, RansomarePoC},
            xticklabel style={xshift=-10pt, rotate=30}],
            legend style={
              at={(0.5,-0.15)},
              anchor=north,
              legend columns=-1
            },
            nodes near coords,
            ymin = 0,
            ymax = 10000,
            every node near coord/.append style={font=\tiny},
            nodes near coords align={vertical},
            ]
        %\addplot coordinates {(Babuk, 10406) (Blackbasta, 10406) (Buhti, 10406)  (Cl0p, 10406) (Conti, 548) (Darkradiation, 10406) (GonnaCry, 106) (JavaRansomare, 2573) (Lockbit, 1) (lollocker, 2858) (monti, 10406) (raasnet, 132) (RansomarePoC, 2304) };
        \addplot coordinates {(Babuk, 10406) (Blackbasta, 10406) (Buhti, 10406)  (Cl0p, 10406) (Conti, 548) (Darkradiation, 10406) (GonnaCry, 106) (JavaRansomare, 2573) (Lockbit, 1) (lollocker, 2858) (monti, 10406) (raasnet, 132) (RansomarePoC, 2304) };
        \addplot coordinates {(Babuk, 0) (Blackbasta, 0) (Buhti, 0) (Cl0p, 68) (Conti, 52)  (Darkradiation, 0) (GonnaCry, 106) (JavaRansomare, 0) (Lockbit, 0) (lollocker, 0) (monti, 68) (raasnet, 0) (RansomarePoC, 481) };
        \legend{Uninterrupted, MTFS Defense}
        
        \end{axis}
        \end{tikzpicture}
    \caption{File Content Encrypted by Various Ransomware Samples}
    \label{fig:my_label}
\end{figure*}
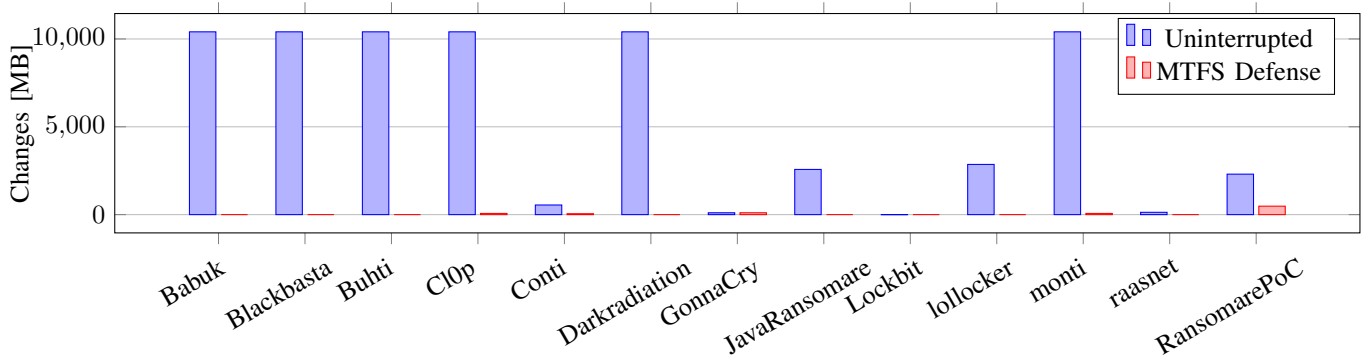

\subsection{Testbed Scenario}
To provide evidence of the effectiveness of \solution{} considering heterogeneous ransomware samples, a more efficient testbed is needed that allows experiments to be executed in a highly observable, reproducible, controllable, and explicit thereby accountable manner. \figurename~\ref{fig:testbed} shows the virtualized testbed developed to execute and defend against ransomware.

In this configuration of the testbed, each experiment takes as input an Ubuntu 22.04 base image, and a set of benign files mounted in the home directory. \cite{garfinkel2009bringing} provides a file corpus of various file types. The files are mounted in a flat file system, where each of the twenty folders in the home directory holds a few hundred megabytes of data. The file corpus is implemented as a ZFS dataset, which enables rapid snapshot creation and comparison later on. To run an experiment, a container of the base image is spawned, the ZFS dataset is mounted, and the ransomware sample is executed for a maximum of five minutes. Five minutes was established as a suitable duration after implementing a simplified ransomware, which encrypts files in a consecutive and single-threaded execution. After the experiment is concluded, a snapshot of the ZFS dataset is created and compared on a per-file basis to the initial state of the dataset. This allows to measure accurately for each of the files when and how it was changed (\ie modified, moved, deleted, or unmodified). For each malware sample, two experiments are  executed. A baseline experiment where the ransomware is executed without interruption, and one where \solution{} is mounted on top of the dataset. For each sample, the number of modified bytes per second is computed.

To gather results, all components of the testbed can be recreated with a single command that creates and provisions a virtual machine. Here, all experiments have been deployed in an Ubuntu 22.04 machine provided by Virtualbox on an Arch Linux host. The host operating system runs an AMD Ryzen 5700G processor running at 4.7 GHz. To ensure that the host storage is not the bottleneck during encryption, the ZFS dataset is executed in a ZFS pool configured as a RAID 1 pool, using two NVMe storage devices (WesternDigital SN550 NVMe SSD). From a file-system perspective, all 20741 files (10460 MB) can be traversed and read in 9 seconds ($\approx$10 Gbit/s) and overwritten in 27 seconds ($\approx$3.2 Gbit/s). 

Finally, the source code and binaries of 24 ransomware families were obtained from open-source platforms and from the malware database \textit{MalwareBazaar}. Each sample is embedded in a script that allows repeatable execution. From the various samples of 24 ransomware families, only 13 samples are runnable in the server, since some samples focus on encrypting volumes from the \textit{ESXi} hypervisor and others require connection to the C\&C server. Three ransomware samples are implemented in the Go Programming language and or as a simple script. Six samples leverage the Python runtime and one runs on the Java Virtual Machine. The remaining eleven samples are available as an Executable and Linkable Format (ELF) binary. This category contains samples involved in recent attacks, such as Hive, Blackbasta, Lockbit, or Clop.

After running each sample against two configurations (\ie uninterrupted and with \solution{}), the results shown in \figurename~\ref{fig:my_label} are obtained. Looking at these results, two aspects can be highlighted. The first and more important conclusion that can be drawn is that for all the samples, \solution{} is able to protect the majority of the data present in the dataset. The worst performance was achieved from \textit{RansomwarePoC} which started to encrypt files before the full set of target files was established. Over all samples considered, 97.01\% of the bytes were saved by deploying the file system. Importantly, it has to be noted that in the testbed scenario outlined, \solution{} operates under conditions where no detection system is available and where files are structured in a flat file system. % where a ransomware can access the first folder, encrypt all files and then move to a subfolder. 
In that sense, it can be concluded that \solution{} can be used both as a honeypot to attract ransomware attacks, while actively mitigating the encryption behavior.
In a second observation, by comparing the datasets before and after deploying the ransomware it can be seen that there are behavioral differences between samples. Most samples differ in terms of encryption rate. Six samples achieved full encryption. Furthermore, the file exploration strategy of different samples varies. Finally, the modification strategy differs since some encrypt the file in place (\ie execution only \texttt{write()} system calls), while others do the same with an additional renaming step. For example, \textit{GonnaCry} adds the \textit{GNNCRY} file extension to highlight encryption. Other ransomware implementations create a new file based on the encrypted content and delete the old file.

\begin{table}[b]
\centering
\caption{Overhead on Benign Workloads}
\label{table:overhead}
\begin{tabular}{@{}lrllll@{}}
\toprule    
            \textit{}  
            & \textit{} &      \multicolumn{2}{c}{\textit{CPU} [\%]} &  \multicolumn{2}{c}{\textit{RAM} [\%]} \\
            
             \textit{Scenario}& \textit{Time} [s] &  \textit{System} & \textit{MTFS} & \textit{System} & \textit{MTFS}
            \\ \midrule
            WL1 & 2.573  & 30.20 & - & 0.5 & -\\
            WL1 + MTFS & 2.576  & 29.48 & 0 &  0.6 & 0.2\\
             \midrule
            WL2 & 36.031 & 90.62 & - & 0.5 & -\\
            WL2 + MTFS & 36.194 & 96.10 & 0 & 0.46 & 0.1 \\
             \midrule
            WL3 & 2.110 & 0.1 & - & 0.3 & -\\
            WL3 + MTFS & 2.017 & 0.1 & 0 & 0.3 & 0.1  \\
             \bottomrule
            
\end{tabular}
\end{table}
\subsection{Overhead and Comparison}
While the defense system allows defense against ransomware in a proactive mode, the performance overhead must be assessed and contrasted. As outlined in \tablename~\ref{table:overhead}, \solution{} was confronted with three workloads -- \texttt{WL1} involves installation of the \textit{Apache Web Server} through a package manager. \texttt{WL2} covers the creation of a \textit{tar} file for archival, and \texttt{WL3} consists of persisting retrieved values from an air pollution sensor over HTTPS. Each workload is executed without \solution{} and secondly by using \solution{} as primary data storage. Most importantly, all workloads can execute uninterrupted, and without additional delays. Furthermore, for all workloads, the overhead occurred by the file system in terms of CPU is negligible while less than 0.2\% of memory are used.

Finally, the defense system can be compared with existing solutions. \cite{icc} presents a comparable approach. Interestingly, proactive defense was found to be ineffective and inefficient against computationally intense malware (\eg ransomware, data leakage). Thus, as shown by the minimal resource consumption while being proactively deployed, \solution{} presents a light-weight defense alternative by relying on a partially virtualized attack surface. Thus, proactive defense based on deception may be more efficient in certain scenarios. When comparing the defense against the same ransomware strain, such a defense may also be more efficient. Although the execution time cannot be compared due to differences in corpus size, in~\cite{icc} 7.1 MB of data is lost, while the same sample did not lead to data loss in \solution{}.

\section{Summary and Future Work}\label{conc}
This paper introduced \solution{}, a file system-based platform which serves as an attack mitigation platform for Moving Target Defense on the file-system abstraction. \solution{} follows the Linux virtual file system architecture to implement an overlay file system, relying on \textit{FUSE} to implement the actual file operations in user space. With that, full control over file operations can be obtained. This allows to implement detection and mitigation strategies in a user space application, enabling tight integration with existing detection approaches that rely on other data sources (\eg hardware counters, performance metrics). Based on this file system, three file-system-based MTD techniques are proposed. %(\texttt{MTD\_INF}, \texttt{MTD\_SUFFIX}, and \texttt{MTD\_DELAY}) are proposed -- one focusing on delaying operations, one presenting a recursive directory graph, and one modifying the magic numbers of files on the fly so that malicious software cannot determine the file type. 
To highlight the feasibility of a file system-based approach, \solution{} was implemented as an overlay file system over existing systems.% using the Go programming language. 

To assess the actual effectiveness, \solution{} and its MTD techniques were evaluated in two heterogeneous scenarios. Experiments performed in the first scenario have shown that the techniques can successfully operate on a Raspberry Pi device and mitigate two ransomware samples. In the second scenario, \solution{} was deployed in a testbed that was specifically created to test ransomware mitigation systems. The testbed enabled execution against many ransomware samples, including 13 samples that were found in real-world malware databases.
The results show that the file-based approach can successfully save a large amount of the data to be protected. %While ransomware is a continuously evolving threat vector, the testbed developed here is the most extensive one in terms of number of samples.

To further develop this work, a ML-based classification system will be investigated. Specifically, it will be analyzed whether it is possible to detect malware-based attacks on the granularity of individual system calls while buffering any modifying operations.
\section*{Acknowledgment}
This work has been partially supported by \textit{(a)} the Swiss Federal Office for Defense Procurement (armasuisse) with the CyberForce project (CYD-C-2020003) and \textit{(b)} the University of Zürich UZH.

\bibliographystyle{IEEEtran}  
\balance
\bibliography{references}

\end{document}